\documentclass[a4paper,onecolumn,11pt,accepted=2023-08-30]{quantumarticle}
\pdfoutput=1
\usepackage[utf8]{inputenc}
\usepackage[english]{babel}
\usepackage[T1]{fontenc}
\usepackage{amsmath}
\usepackage{hyperref}

\usepackage{tikz}
\usepackage{lipsum}

\usepackage{amsfonts}
\usepackage{amsthm}
\usepackage{amscd}
\usepackage{bbold}
\usepackage{t1enc}
\usepackage[mathscr]{eucal}
\usepackage{indentfirst}
\usepackage{graphicx}
\usepackage{graphics}
\usepackage{pict2e}
\usepackage{xcolor}
\theoremstyle{plain}
\newtheorem{Th}{Theorem}%[section]
\newtheorem{Lemma}[Th]{Lemma}
\newtheorem{Cor}[Th]{Corollary}

\newtheorem{Pro}[Th]{Proposition}

 \theoremstyle{definition}

\newtheorem{Rem}[Th]{Remark}
\newtheorem{?}[Th]{Problem}
\DeclareMathOperator{\Res}{Res}
\DeclareMathOperator{\diag}{diag}

\DeclareMathOperator{\sgn}{sgn}

\newcommand{\R}{\mathbb{R}}

\newcommand{\C}{\mathbb{C}}
\newcommand{\im}{\operatorname{im}}
\newcommand{\tr}{\operatorname{tr}}

\begin{document}

\title[Integral formula  for quantum relative entropy]
{Integral formula  for quantum relative entropy implies  data processing inequality}

\author{P\'eter E.\ Frenkel}

\address{E\"{o}tv\"{o}s  University, Institute of Mathematics,
  P\'{a}zm\'{a}ny P\'{e}ter s\'{e}t\'{a}ny 1/C, Budapest, 1117 Hungary \\ and R\'enyi Institute,  Budapest, Re\'altanoda u.\ 13-15, 1053 Hungary}
\email{frenkelp265@gmail.com}

\thanks{This research was partially supported by MTA R\'enyi
``Lend\"ulet'' Groups and Graphs Research Group and by NKFIH  grants K~124152 and KKP 139502.
}

%This paper was presented in part at the School on Advanced Topics in Quantum Information and Foundations, February 2021.
%This work has been submitted to the IEEE for possible publication. Copyright may be transferred without notice, after which this version may no longer be accessible.}

 %\subjclass[2020]{}


\begin{abstract}Integral representations of quantum relative entropy, and of  the  directional second and higher order  derivatives of von Neumann entropy, are established, and used to give  simple proofs of   fundamental,  known data processing inequalities: the Holevo bound on the quantity of information transmitted by a quantum communication channel, and, much more generally, the monotonicity  of quantum relative entropy under trace-preserving positive linear maps --- complete positivity of the map need not be assumed. The latter result was first proved by  M\"uller-Hermes and Reeb,  based on work of Beigi.%In fact, a more general known result,  monotonicity of the Holevo quantity under positive trace-preserving linear maps, is recovered.

For a simple  application of such  monotonicities, we consider any  `divergence'
 that is non-increasing under  quantum measurements, such as    the concavity of von Neumann entropy, or various known quantum divergences.
An elegant argument due to Hiai,  Ohya, and  Tsukada is used  to show that the infimum of such a `divergence'  on pairs of quantum states with prescribed trace distance is the same as the corresponding infimum on  pairs of binary classical states.

Applications of the new integral formulae to the general probabilistic  model of information theory, and  a related integral formula for the classical R\'enyi divergence, are also discussed.
\end{abstract}

\maketitle

\section{Introduction}
Half a  century ago, Alexander Holevo  proved  his famous inequality:  the quantity of information transmitted  by a quantum communication channel using a  given ensemble of quantum states is bounded from above by the extent to which  von Neumann entropy is concave on the  ensemble. One of the main ingredients in Holevo's proof is  an explicit, closed formula for the  directional second derivative $S''$ of the von Neumann entropy.

Since that time, the Holevo bound has become an important building block  of the vast theory of quantum information. Generalizations and alternative proofs, often using advanced methods of that theory, have been given.
Almost simultaneously with Holevo's work, Elliott Lieb and Mary Beth Ruskai~\cite{LR} established the strong subadditivity of von Neumann entropy, which quickly led to G\"oran Lindblad's proof~\cite{L} of monotonicity of quantum relative entropy under completely positive trace-preserving linear maps --- a generalization of Holevo's inequality.
Much later, a further generalization was proved by A.\ M\"uller-Hermes and D.\ Reeb~\cite{MR}, based on work of S.\ Beigi~\cite{B}: quantum relative entropy cannot increase under  a trace-preserving positive linear map --- complete positivity of the map need not be assumed.

 In this paper,
we return to more classical methods of analysis and linear algebra. In Section~\ref{neumann}, we   prove the alternative  formula \begin{equation*}
-\left.\frac1{2} S(\rho+t\sigma)''\right\vert_{t=0}=\int_{-\infty}^{\infty}\frac{dt}
{|t|^{3}}\tr ^ -(\rho+t\sigma)
\end{equation*} for the  directional second derivative of von Neumann entropy, which then leads to similar formulas  for  directional derivatives of higher order.   Note that $\tr^-$ stands for the sum of absolute values of negative eigenvalues.

 Before that, in Section~\ref{qre}, we establish a  similar formula  for the quantum relative entropy.  The simplest form of this formula is \begin{align}\label{relentformula}
D(\rho\|\sigma)=\int_{-\infty}^{\infty}\frac{dt}
{|t|(t-1)^{2}}\tr ^ -((1-t)\rho+t\sigma),
\end{align}  which holds for any two quantum states.

We then show in Section~\ref{dataprocess} that our formulae  lead to the above mentioned monotonicity of quantum relative entropy and to the Holevo inequality in a very simple way, and a characterization of the case of equality in these inequalities can be deduced from our new proof.  %In fact, a more general known result,  monotonicity of the Holevo quantity under positive trace-preserving linear maps, is recovered.
In a recent preprint~\cite{J}, it  is pointed out that the $\tr^-$ appearing in~\eqref{relentformula} is closely related to optimal error probabilities in quantum state discrimination, and, as a consequence,
\eqref{relentformula} leads to a new characterization of recoverability of quantum states with respect to a quantum channel in terms of quantities related to quantum hypothesis testing.

In Section~\ref{lowerbounds},  we  consider any  `divergence-like' quantity that is non-increas\-ing under  quantum measurements, such as   the concavity of von Neumann entropy, sandwiched R\'enyi divergences, or the more general optimized quantum $f$-divergences. We use a  simple  construction  due to Hiai,  Ohya, and  Tsukada~\cite{HOC}  from the year 1981 to show that the infimum of such a  `divergence' on pairs  with prescribed trace distance is the same for (arbitrary dimensional)  quantum states  as for binary classical states.
For example,  the Holevo inequality is used to obtain  a new, tight lower bound on the concavity of von Neumann entropy, improving on a lower bound given by Isaac Kim in 2014.

In Section~\ref{GPT}, we discuss how formula~\eqref{relentformula} can be applied in the general probabilistic framework of information theory, yielding, in particular, an extension of the Holevo inequality to that framework.
In Section~\ref{renyi}, we provide an integral representation of the classical R\'enyi relative entropy and discuss its possible relevance to  quantum information theory and to general probabilistic theory.

\section*{Acknowledgements} %Data sharing is not applicable to this article as no datasets were generated or analysed during the current study.
The author is grateful to Mil\'an Mosonyi, P\'eter Vrana, and two unnamed referees for helpful comments. %has no competing interests to declare that are relevant to the content of this article.

 \section{Notations, terminology, and basic trace inequalities}
\rm %The set $\{1,\dots, k\}$ is denoted by  $[k]$.
 We write $\log$ for the natural logarithm.
 Partial derivatives will be denoted by putting the corresponding variable in the  subscript. A $'$ means differentiation with respect to $t$.

  The set of $n$-square matrices with complex entries is written $M_n(\C)$. The identity matrix is $\bf 1$.
A complex  matrix $A$  is \it psdh \rm if it is positive semi-definite Hermitian, written $A\ge 0$. We write $A\ge B$ to mean that $A-B\ge 0$.  %Positive definiteness is denoted by $A>0$.
   % For a real number $a$, we write  $a^+=\max (a, 0)$ and  $a^-=-\min (a, 0)=(-a)^+$.
 For a Hermitian matrix $A$, we write $A=A^+-A^-$, where $A^\pm\ge 0$ and $A^+A^-=A^-A^+=0$. We define $|A|=A^++A^-$. We write $\tr^\pm A=\tr A^\pm$  and $\|A\|_1=\tr |A|$ for the sum of absolute values of positive/negative eigenvalues and all eigenvalues of $A$, respectively.  %A \emph{positive operator valued measure (POVM)} is a sequence $E_1$, \dots, $E_k$ of psdh matrices summing to $\bf 1$.
  Recall that \[\tr^+A=\max\{\tr PA: P^2=P=P^*\}.\] The maximum is attained if and only if \begin{equation}\label{projmax}({\bf 1}-P)A^+=PA^-=0.\end{equation} If $A\ge B$, then $PAP\ge PBP$ and therefore $\tr PA\ge \tr PB$ for all projections $P$, whence \begin{equation}\label{basic}\tr^+A\ge\tr^+ B.\end{equation} There is equality here if and only if  there exists a projection $P$ satisfying~\eqref{projmax} and $\tr PA=\tr PB$. In this case, $\tr PA^+=\tr PA+\tr PA^-=\tr PA=\tr PB$, but $\im A^+\perp\im ({\bf 1}-P)=\ker P$, whence $$\im A^+\subseteq\im P\perp \im (A^+-B),$$ so $A^+(A^+-B)=0$.
Conversely, if this last equality holds, then the projection $P$  onto $\im A^+$ satisfies \eqref{projmax} and also $$0\le \tr P(A-B)\le\tr P(A^+-B)=0,$$ so there is equality in~\eqref{basic}.

   Recall the well-known fact that \begin{equation}\label{convex}\tr^\pm {\text {are nonnegative  convex functions.}}\end{equation} Indeed, from~\eqref{basic}, the inequality $$\tr^+(A+B)\le\tr^+(A^++B^+)=\tr(A^++B^+)=\tr A^++\tr B^+=\tr^+A+\tr^+B$$ holds  for any self-adjoint $A$ and $B$, and the identity $\tr^+tA=t\tr^+A$  holds  for any $t\ge 0$ and self-adjoint $A$, implying convexity of $\tr^+$. Since $\tr^-(A)=\tr^+(-A)$, the function $\tr^-$ is also  convex.

A \emph{density matrix}, or \emph{quantum state},  is a psdh matrix with trace 1.  The  \emph{von Neumann entropy} of a psdh matrix $\rho$ (of arbitrary trace) is $S(\rho)=-\tr\rho\log\rho$.
The \emph{quantum relative entropy} (Umegaki~\cite{U}) of two psdh matrices $\rho$ and $\sigma$  (of arbitrary trace) is $$D(\rho\|\sigma)=\begin{cases}\tr\rho(\log\rho-\log\sigma) \textrm{ if }\im\rho\subseteq\im\sigma\\  +\infty\textrm{  otherwise.}\end{cases}$$

%In the first two subsections, w
\section%{Preliminaries}%We introduce a few basic tools.
%\subsection{Derivative of matrix logarithm}  Let $X(t)$ be a differentiable curve  whose values are  positive definite matrices.
%By \cite[formula (7)]{A}, we have $$(\log X)'=\int_0^{\infty}(X+r{\bf 1})^{-1}X'(X+r{\bf 1})^{-1}\; dr.$$
%From this, we infer
%\begin{Lemma}\label{trcomm}  If, for a given $t$, $X(t)$ commutes with a matrix $Y$, then $$\tr Y(\log X)'(t)=\tr YX'(t)X(t)^{-1}.$$
%\end{Lemma}
%\subsection
{Eigenvalues of matrix pencils}\label{pencil}
It will be useful to study the negative real eigenvalues of the linear  matrix pencil
\begin{equation}\label{pencildef}A(t)=(1-t)\rho+t\sigma,\end{equation} %\quad\textrm{ or }\quad A(t)=\rho+t\sigma,$$
  where $A(0)=\rho\ge 0$ and $A(1)=\sigma$ is Hermitian. A priori, we allow $t\in\C $  here. Thus, in general, $A(t)$ is not  Hermitian, but some real eigenvalues could occur for some non-real $t$. However, we show that \emph{negative} real eigenvalues of $A(t)$
 can only occur for real $t$.
 \begin{Lemma}\label{real}
  If $A(t)e=-re$ for a  unit vector $e$ and a positive real number $r$, then $t((\rho-\sigma)e, e)>0$, and therefore $t\ne 0$ is real.
%  Therefore, we have
 \end{Lemma}
  \begin{proof}We have  $-r=(A(t)e,e)=(1-t)(\rho e, e)+t(\sigma e, e)$, whence $$t((\rho-\sigma)e, e)=(\rho e, e)+r>0.$$  Thus, $t\ne 0$.  Since $\rho$ and $\sigma$ are both Hermitian, $((\rho-\sigma)e, e)$ is real and therefore so is $t$.
  \end{proof}
   %Since Hermitian matrices with multiple eigenvalues have codimension $>1$, we will be able to

   Consider the two-parameter family of matrices \begin{equation}\label{X}X(t,r)=A(t)+r\bf 1.\end{equation} Its partial derivatives are $X_t=\sigma-\rho$ and $X_r=\bf 1$.

 Define the bivariate polynomial \begin{equation}\label{f}f(t,r)=\det {\color{black}{X(t,r)}}%=\prod_{m=1}^n(r_m(t)+r)
 .\end{equation} For its partial logarithmic derivatives, we have, by Jacobi's formula,  \begin{equation}\label{Jacobi}f_t/f=\tr (\sigma-\rho)X^{-1}\qquad {\text {and} }\qquad  f_r/f=\tr X^{-1}\end{equation} whenever $X$ is invertible, i.e., whenever $f\ne 0$. In this case, \begin{equation}\label{ratio}f_t\tr X^{-1}=f_r\tr(\sigma-\rho)X^{-1}.\end{equation}

 We have $f(t,r)=0$ if and only if $-r$ is an eigenvalue of $A(t)$. In this case, the ratio of partial derivatives is given by
 \begin{Lemma}Let $t$ be real. If $A(t)e=-re$ for  a  unit vector $e$,  then  $${f_t}(t, r)=((\sigma-\rho) e, e)f_r(t,r).$$
 \end{Lemma}
 \begin{proof}%The  equality is easily seen after including $e$ in an orthonormal eigenbasis of
 Let $P$ be the  orthogonal  projection to the $(-r)$-eigenspace of $A(t)$. We have $A(t)=-rP+({\bf 1}-P)A(t)({\bf 1}-P)$, where the second term does not have $-r$  as an eigenvalue. {\color{black} Thus, for all $s$, we have $X(t,s)=A(t)+s{\bf 1}=(s-r)P+({\bf 1}-P)(A(t)+s{\bf 1})({\bf 1}-P)$, where the second term does not have $s-r$  as an eigenvalue. } Then $X(t,s)^{-1}\sim P/(s-r)$ as $s\to r$. Thus, \eqref{ratio} yields $f_t(t,r)\tr P=f_r(t,r)\tr(\sigma-\rho)P$.  If $-r$ is a simple eigenvalue of $A(t)$, then $\tr P=1$ and $\tr(\sigma-\rho)P=((\sigma-\rho)e, e)$, proving the Lemma.  If $-r$ is a multiple eigenvalue of $A(t)$, then $f_r(t,r)=0$, {\color{black}  but $\tr P\ne 0$,} hence $f_t(t,r)=0$, and the Lemma holds in a trivial way.
 \end{proof}
 These two lemmas  imply that any negative simple eigenvalue of   $A(t)$   gets more negative as $t$ moves farther away from zero.  More precisely, we have
 \begin{Cor}\label{kulcs} If $f(t,r)=0$ and  $r>0$, then
\begin{enumerate}
 \item[(a)] $t \ne 0$ is real, and $tf_r(t,r)f_t(t,r)\le 0$.

\item[(b)] The following are equivalent: $f_t(t,r)=0$; $f_r(t,r)=0$; $-r$ is a multiple eigenvalue of $A(t)$.
\end{enumerate}
 \end{Cor}
We have $f(t,r)=f_r(t,r)=0$ if and only if $-r$ is a multiple eigenvalue of $A(t)$.  As a  polynomial in $r$, $f$ has  a discriminant whose value is a polynomial in $t$. The discriminant is zero for a given value of $t$ if and only if the matrix $A(t)$   has a  multiple eigenvalue. This happens either for finitely many $t$ or for all $t$.  %Call these values of $t$ bad and other values of $t$ good.  Assume that there is at least one  good  $t$. %, the matrix $A(t)$   has simple eigenvalues only.

   \section{Quantum relative entropy}\label{qre}  Let $\rho$ and $\sigma$ be psdh matrices. The notations~\eqref{pencildef}, \eqref{X}, and  %s , in particular
  \eqref{f} introduced in Section~\ref{pencil} will be used.
   We wish to prove an integral formula for the quantum relative entropy $D(\rho\|\sigma)$. In the first part of the proof, we  join the pair $(\rho, \sigma)$  to infinity in the direction of the identity matrix. This is done in the following two lemmas, and yields an integral formula in terms of certain logarithmic derivatives of the function $f(t,r)$. We will then  convert this into an integral  formula in terms of  $\tr^- A(t)$ by applying the Residue Theorem and by changing the integration variable from $r$ to $t$.
 \begin{Lemma}\label{limit}$\lim_{r\to\infty} D(\rho+r{\bf 1}\|\sigma+r\bf 1)=\tr (\rho-\sigma)$.
 \end{Lemma}

 \begin{proof} $D(\rho+r{\bf 1}\|\sigma+r{\bf 1})=rD({\bf 1}+\rho/r\|{\bf 1}+\sigma/r)\sim r\tr(\rho-\sigma)/r.$
 \end{proof}

 %We set $A(t)=(1-t)\rho+t\sigma$ for this section.

 \begin{Lemma}\label{derivative}\begin{enumerate}
 \item[(a)] For all $r>0$, we have \begin{align*}&\frac d{dr} D(\rho+r{\bf 1}\|\sigma+r{\bf 1})=\\=\tr\log(\rho+r{\bf 1})& -\tr\log(\sigma+r{\bf 1})+\tr(\sigma-\rho)(\sigma+r{\bf 1})^{-1}
 %\log A(t)+r{\bf 1} )'(1,r)
 =\\=\log f(0,r)&-\log f(1, r)+(\log f)'(1,r)\end{align*}

     \item[(b)] If $\im\rho\subseteq\im\sigma$, then the expression above is $o(1/r)$
   as $r\to 0$ or $r\to\infty$.
   \end{enumerate}
 \end{Lemma}

 \begin{proof} (a) On the left hand side, we differentiate  the product in the argument of $\tr$ to get $$\tr \left(\log(\rho+r{\bf 1})- \log (\sigma+r{\bf 1})+(\rho+r{\bf 1})\left((\rho+{r\bf 1})^{-1}-(\sigma+{r\bf 1})^{-1}\right)\right).$$
 Apply the identity  $${\bf 1}-(\rho+r{\bf 1})
(\sigma+r{\bf 1})^{-1}=(\sigma-\rho)(\sigma+r{\bf 1})^{-1}$$   to arrive at the middle expression in Statement (a) of the Lemma. By $\tr\log=\log\det$, the first two terms are clearly equal to the first two terms of the last expression in (a). By~\eqref{Jacobi}, the third terms are also equal. %In the third term, observe that $\sigma-\rho=A'(1)$ and  $\sigma=A(1)$. Jacobi's formula $(\det X)'=\det X\cdot\tr X'X^{-1}$ can be applied to the pencil $X(t)=A(t)+r\bf 1$ to yield}}
%$$\tr A'(1)(A(1)+r{\bf 1})^{-1}=(\log\det (A+r{\bf 1}))'(1),$$  {\color{black}{which is the third term  of the last expression in (a).}}%Once again, apply}} $\tr\log=\log\det$ to get to the bottom line.

(b) When $r\to 0$, the first two terms are $O(\log(1/r))$, and the third term   $\to \tr(\sigma-\rho)\sigma^{(-1)}$ if $\im\rho\subseteq\im\sigma$, where $\sigma^{(-1)}=(\sigma\vert_{\im\sigma})^{-1}\oplus 0$ is the pseudoinverse of $\sigma$. %, i.e. the direct sum of the inverse of the restriction of $\sigma$ to $\im\sigma$  and the zero operator on $\ker\sigma$.

When $r\to\infty$, the third term is $\sim\tr(\sigma-\rho)/r$, and  the sum of the first two terms is $\tr\log({\bf 1}+\rho/r)-\tr\log({\bf 1}+\sigma/r)\sim \tr(\rho-\sigma)/r$. %, may be as large as $O(1/r)$, but the sum of three terms is $O(1/r^2)$.
 \end{proof}

 From Lemmas~\ref{limit} and~\ref{derivative}(a), we have
 \begin{align*}
 D(\rho\|\sigma)+\tr(\sigma-\rho)
 =
 -\int_0^\infty
 \left(
 \log f(0,r)-\log f(1, r)+(\log f)'(1,r)
 \right)
 \;dr.
  \end{align*} Now we integrate by parts. If $\im\rho\subseteq \im\sigma$, then, by Lemma~\ref{derivative}(b), the  result is simply $$
 \int_0^\infty r\cdot
  \left(
  h(0,r)-h(1, r)+h'(1,r)
  \right)
  \;dr, $$  where $h=(\log f)_r=f_r/f$ is a rational function of  $t$ and  $r$, holomorphic unless $f=0$, so certainly holomorphic at $(t,r)$  when $r>0$ and $t\in [0,1]$. Therefore, as a rational function of $t$, the function  $h/t$ has residue $h(0,r)$ at 0 and is holomorphic at 1, the function $h/(t-1)$ has residue $h(1,r)$ at 1 and is holomorphic at 0,  and the function $h/(t-1)^2$ has residue $h'(1,r)$ at $1$ and is holomorphic at $0$.
   Let $$g(t)=\frac1t-\frac1{t-1}+\frac1{(t-1)^2}=\frac1{t(t-1)^2},$$  then \begin{align}\label{lastintegral}
 D(\rho\|\sigma)+\tr(\sigma-\rho)
 =
  %=\\=
 \int_0^\infty
  r\cdot(\Res_{t=0}+\Res_{t=1})
(  g{\color{black}{h}})%(0,r)-(\log f)_r(1, r)+(\log f)_r'(1,r)
  \;dr.
 \end{align}
  For a  fixed $r> 0$, observe that
$g{\color{black}{h}}$, %(with no differentiation in the denominator),
 as  a rational function of $t$  in the complex plane, is holomorphic except where $t=0$, $t=1$, or $f=0$. The latter case occurs if and only if $-r$ is an eigenvalue of $A(t)$. If $-r$ is a \emph{simple}  eigenvalue of $A(t)$, then $f_t(t,r)\ne 0$ by Corollary~\ref{kulcs}, and the residue of ${\color{black}{gh=}}gf_r/f$ at $t$ is
  $ g{f_r}/f_t$ {\color{black}{because $gf_r$ is holomorphic at $t$ and  $f(t,r)=0$.

 {\color{black} The bivariate polynomial $f$, when viewed as  a polynomial in $t$, has a leading coefficient that is a univariate polynomial in $r$.}   For all but finitely many  {\color{black}  values of $r$,  this leading coefficient is nonzero, and then} we have \begin{equation}\label{bounded}h= O\left(1\right)\qquad
     ({\color{black}|}t{\color{black}|}\to\infty),\end{equation} whence $ gh =O(|g|)= O\left(|t|^{-3}\right)$,}} so  the contour integrals on circles $|t|=T$  tend to zero as $T\to\infty$.  By the Residue Theorem,  the sum of all residues {\color{black}{of $gh$}} is zero.

 {\color{black}{   Let us assume that not all  $A(t)$ have multiple eigenvalues. In this case, only finitely many $A(t)$ have multiple eigenvalues, therefore only finitely many numbers $-r$ occur as multiple eigenvalues of some $A(t)$. These finitely many values of $r$, together with the ones for which \eqref{bounded} fails, do not influence the integral in~\eqref{lastintegral}.}}  The right hand side of \eqref{lastintegral} therefore becomes \begin{equation}\label{r-integral}-\int_0^{\infty}r
\sum_{f(t,r)=0}\frac{gf_r}{f_t}\; dr.\end{equation}
{\color{black}{By Lemma~\ref{real}, only real numbers $t$ can satisfy  the condition $f(t,r)=0$  required in the summation.
Therefore, we can think of~\eqref{r-integral} as an integral on the portion of the real  algebraic plane curve $f(t,r)=0$ that lies  in the upper half-plane $r>0$.   The finitely many singular points that the curve may have, corresponding to $-r$ being a  multiple eigenvalue of $A(t)$, do not influence the integral. We only need to study the integral along the smooth arcs of the curve. Each smooth arc is parametrized by the variable $r$, but we want to reparametrize it using the variable $t$. For  a simple negative eigenvalue $-r$ of $A(t)$, Corollary~\ref{kulcs} implies that   $f_r(t,r)\ne 0$ and $|dr/dt|=|f_t/f_r|=-(\sgn t)f_t/f_r$ as we move along  the  curve. Hence,  by the rule for change of  variables, the integral is rewritten as}} $$\int_{-\infty}^{\infty}\frac {dt}
{|t|(t-1)^2}\sum {\color{black}{\left(r\;:\; r>0,\;f(t,r)=0\right)}}.$$  {\color{black} The  change of variables is justified since the integrand is positive at every smooth point.} % because of Lemmas~\ref{real} and \ref{monoton},
%{\color{black}{This equality holds }} since, f $f=0$ {\color{black}{that are located in the upper half-plane}}.

 The last sum that has appeared is $\tr^- A(t)$.  We {\color{black} arrive  at the main result of this paper.}
\begin{Th}\label{relent}
Let $\rho, \sigma\in M_n(\C)$ be psdh matrices. %, where $\rho$ is a density matrix, $\sigma$ is Hermitian with trace zero, and $\im\sigma\subseteq\im\rho$.
Then
\begin{align*}%\label{relentformula}
D(\rho\|\sigma)=\tr(\rho-\sigma)+\int_{-\infty}^{\infty}\frac{dt}
{|t|(t-1)^{2}}\tr ^ -A(t)=\\=\int_{-\infty}^{0}\frac{dt}
{|t|(|t|+1)^{2}}(\tr ^ +A(t)-\tr\rho)+\int_{1}^{\infty}\frac{dt}
{t(t-1)^{2}}\tr ^ -A(t)
,
\end{align*}
where $A(t)=(1-t)\rho+t\sigma$, and $\tr^\pm$ stands for the sum of absolute values of positive and negative eigenvalues, respectively.
\end{Th}

 \begin{proof} First equality: Both sides are $+\infty$ unless $\im\rho\subseteq\im\sigma$, which we henceforth assume. Restricting our attention to the image of $\sigma$, we may assume that $\sigma$ is positive definite to begin with.
 Since both sides are then continuous, we may change  $\sigma$ a little bit so that it has no  multiple eigenvalues.
 %Since Hermitian matrices with multiple eigenvalues have codimension $>1$, we may assume that $(1-t)\rho+t\sigma$  has simple eigenvalues  $r_1(t)<\dots <r_n(t)$ for all $t$.  These depend smoothly on $t$ by the Implicit Function Theorem. Define \begin{equation}\label{f}f(t,r)=\det((1-t)\rho+t\sigma+r{\bf 1})=\prod_{m=1}^n(r_m(t)+r).\end{equation}
The preceding discussion then applies and  the first equality is proved.

Second equality: For $0\le t\le 1$, we have $A(t)\ge 0$ and therefore $\tr^- A(t)=0$. For $t<0$, observe that \begin{equation}\label{atiras}\tr ^ +A(t)-\tr\rho=\tr ^ -A(t)+|t|\tr(\rho-\sigma).\end{equation} Then use $\int_{-\infty}^0dt/(t-1)^2=1$ to conclude.
%then applies with $$A(t)=(1-t)\rho+t\sigma.$$
% The integral on the right hand side of \eqref{relentformula} can be rewritten as \begin{equation}\label{relenthelyettes}\int_{-\infty}^{\infty}\frac {dt}
%{|t|(t-1)^2}\sum_{f(t,r)=0}r^+\: dt=-\int_0^{\infty}r
%\sum_{f(t,r)=0}\frac{f_r}{t(t-1)^2f_t}\; dr.\end{equation} The summand here can be interpreted as  a residue in the  following way. For a  fixed $r\ge 0$, observe that
%$ {f_r}/({t(t-1)^2f})$ (with no differentiation in the denominator), as  a rational function of $t$  in the complex plane, is holomorphic except where $t=0$, $t=1$ or $f=0$. In the latter case, the residue is
%  $ {f_r}/({t(t-1)^2f_t})$. By the Residue Theorem,  the sum of all residues is zero because the function is
%    $O\left(|t|^{-3}\right)$
%    as $t\to\infty$, so  the contour integrals on circles $|t|=T$  tend to zero as $T\to\infty$.
 %   The residue at $t=0$ is $$(f_r/(t-1)^2f)(0,r)=\sum_{m=1}^n\frac1{r_m(0)+r}.$$    The residue at $t=1$ is $$(f_r/tf)'(1,r)=-\sum_{m=1}^n\left(\frac1{r_m(1)+r}+\frac{r_m'(1)}{(r_m(1)+r)^2}\right).$$ EDDIG JO
 % Therefore,  the right hand side of \eqref{relenthelyettes}  is \begin{equation}\label{relentintegral}\int_0^\infty r(\log f)_r''(0,r)\; dr.\end{equation}
\end{proof}

From Theorem~\ref{relent} {\color{black}{and fact~\eqref{convex},}} we immediately recover the well-known fact that $D(\rho\|\sigma)$ is a convex function of the pair $(\rho, \sigma)$, and it is nonnegative whenever $\tr\rho\ge\tr\sigma$.
%\subsection{First  directional derivative of von Neumann entropy}
\section{Higher order derivatives of von Neumann entropy}\label{neumann}
In this section, $\rho$ is a psdh %density
 matrix and  $\sigma$ is a %traceless
  Hermitian matrix with $\im\sigma\subseteq\im\rho$. %, and $A(t)=\rho+t\sigma$.
We wish to find an integral formula for $S{\color{black}{\left.(\rho+t\sigma)^{(m)}\right\vert_{t=0}}}$ when $m\ge 2$.
When $m=2$, $\tr\rho=1$, and $\tr\sigma=0$, an explicit formula for this quantity, in terms of the spectral decomposition of $\rho$, has been given by A.\ S.\ Holevo  in his seminal paper \cite[Lemma 4]{H}.
%We give an alternative  formula for it.
The fact that our integral formula yields  the same value seems non-obvious. The  proof given  below  does not rely on Holevo's explicit formula.

\begin{Th}\label{main}
Let $\rho, \sigma\in M_n(\C)$ with $\rho\ge 0$,
 $\sigma^*=\sigma$,
 and $\im\sigma\subseteq\im\rho$.
\begin{enumerate}
\item[(a)]   For all $m\ge 2$,  we have \begin{equation}\label{formula}
-{\color{black}{\left.\frac1{m!} S(\rho+t\sigma)^{(m)}\right\vert_{t=0}}}=\int_{-\infty}^{\infty}\frac{dt}
{|t|t^{m}}\tr ^ -(\rho+t\sigma),
\end{equation}
where $\tr^-$ stands for the sum of absolute values of negative eigenvalues.

\item[(b)] When $m\ge 2$ is even, the quantity \eqref{formula} is nonnegative and convex as  a function of the pair $(\rho, \sigma)$.
\end{enumerate}
\end{Th}

\begin{proof}(a) Case $m=2$: We have $$- {\color{black}{\left.S(\rho+t\sigma)''\right\vert_{t=0}}}=\lim_{t\to 0}\frac1{t^2}\left(2S(\rho)-S(\rho+t\sigma)-S(\rho-t\sigma)\right).$$
In the parentheses here, we have $$D(\rho+t\sigma\|\rho)+D(\rho-t\sigma\|\rho).$$  By Theorem~\ref{relent}, we have $$D(\rho\pm t\sigma\|\rho)\mp t\tr\sigma=\int_{-\infty}^\infty\frac{ds}{|s|(s-1)^2}\tr^-(\rho\pm(1-s)t\sigma).$$   Putting $u=(1-s)t$, i.e.\ substituting $s=1-u/t$, this becomes $$t^2\int_{-\infty}^\infty\frac{du}{|t-u|u^2}\tr^-(\rho\pm u\sigma)\sim t^2\int_{-\infty}^\infty\frac{du}{|u|^3}\tr^-(\rho+ u\sigma)$$ as $t\to 0$ by Lebesgue's Dominated Convergence Theorem.
Case $m=2$ follows.

 If the statement holds for $m$, then
\begin{align*}-{\color{black}{\left.\frac1{m!} S(\rho+t\sigma)^{(m)}\right\vert_{t=u}=-\left.\frac1{m!} S(\rho+u\sigma+t\sigma)^{(m)}\right\vert_{t=0}}}=\\=\int_{-\infty}^{\infty}\frac{dt}
{|t|t^{m}}\tr ^ -(\rho+(t+u)\sigma)=\int_{-\infty}^{\infty}\frac{dt}
{|t-u|(t-u)^{m}}\tr ^ -(\rho+t\sigma) \end{align*}for {\color{black}{ $|u|<T$, where $T>0$ is such that $\rho\pm  T\sigma\ge 0$ {\color{black} for both signs}.

When  $|u|<|t|$, we have \[\frac d{du}\frac1{|t-u|(t-u)^m}=\frac{m+1}{|t-u|(t-u)^{m+1}},\]  whose absolute value is bounded from above by  \[\frac{(m+1)2^{m+2}}{|t|^{m+2}}\] if $|u|\le |t|/2$.  Thus, for $|u|\le T/2$, the absolute value of the derivative with respect to $u$ of the last integrand is dominated by \[\frac{(m+1)2^{m+2}}{|t|^{m+2}}\tr ^ -(\rho+t\sigma),\] which is integrable on the real line.}} Use Lebesgue's Theorem to differentiate w.r.t.\ $u$ at 0 under the integral sign, and get
 Theorem~\ref{main}(a) for $m+1$.

 (b)   {\color{black}{From~\eqref{convex}, we see}} that $\tr^-(\rho+t\sigma)$ is nonnegative  and convex as  a function of the pair $(\rho, \sigma)$.  For $m$ even, $|t|t^m\ge 0$ for all $t$.
\end{proof}

\section{Data processing inequalities}\label{dataprocess}
Let $\mathcal E:M_n(\C)\to M_{n'}(\C)$ be a trace-nonincreasing positive linear map. \emph{Positivity} means that psdh matrices are {\color{black}{mapped}} to psdh matrices (and therefore Hermitian matrices are {\color{black}{mapped}} to Hermitian matrices). \emph{Trace-nonincreas\-ing} means that $\tr\mathcal E A\le\tr A$ for all $A\ge 0$. % and density matrices are taken to density matrices.
An important example %of a positive trace-preserving  linear map
 is given by  a  \emph{positive operator valued measure}, or \emph{partition of unity}: psdh matrices $E_1$, \dots, $E_k$ summing to ${\bf 1}$, which  give rise to a completely positive, trace-preserving  linear map, the \emph{quantum measurement} $$\mathcal E:A\mapsto\diag( \tr E_1A,  \dots, \tr E_kA).$$

\begin{Lemma}\label{proc}\begin{enumerate}
\item[(a)] For any  trace-nonincreasing positive  linear map $\mathcal E$ and any Hermitian  matrix  $A$, we have $\tr^{\pm}\mathcal EA\le \tr^{\pm}A$.
\item[(b)] Equality holds in the statement (a) if and only if $\tr\mathcal E A^\pm=\tr A^\pm$ and  $\mathcal EA^+\mathcal EA^-=0$.
\item[(c)]For a quantum measurement, the condition of equality in (a) is that for all $i$, we have    $ E_iA^+=0$ or  $ E_iA^-=0$.
\end{enumerate}
\end{Lemma}

\begin{proof}%Since $\tr^+\mathcal EA-\tr^-\mathcal EA=\tr\mathcal EA=\tr A=\tr^+A-\tr^-A,$ i
It suffices to treat the $+$ case because passing from $A$ to $-A$ interchanges $\tr^+$ and $\tr^-$ as well as $A^+$ and $A^-$.

(a) {\color{black}{We have $A\le A^+$ and therefore $\mathcal EA\le\mathcal EA^+$. From~\eqref{basic}, we get that}}
\begin{align*}
\tr^+\mathcal EA%=\tr^+\mathcal E (A^+-A^-)=\tr^+(\mathcal  EA^+-\mathcal EA^-)\le \\
\le \tr^+\mathcal  EA^+=\tr\mathcal  EA^+
\le\tr A^+=\tr^+A.\end{align*}

{\color{black}{(b) Since  $(\mathcal EA^+)^+=\mathcal EA^+$ and  $\mathcal EA^+-\mathcal EA=\mathcal EA^-$, the claim follows from the condition for equality in~\eqref{basic}.}}

(c) For a quantum measurement, the condition of equality is that there be no $i$  with $\tr E_iA^\pm>0$ for both signs. \end{proof}

{\color{black} {
\begin{Rem}\label{equivalence} The condition  $\mathcal EA^+\mathcal EA^-=0$ appearing in (b) is equivalent to each of the following: $(\mathcal EA)^+=\mathcal EA^+$, $(\mathcal EA)^-=\mathcal EA^-$,  $\|\mathcal EA\|_1=\tr \mathcal E|A|$.
When  $\tr \mathcal E|A|=\tr |A|$, % is trace-preserving,
 it is also equivalent to $\|\mathcal EA\|_1=\|A\|_1$.
 \end{Rem}}}

\subsection{Quantum relative entropy}
From Theorem~\ref{relent} and  Lemma~\ref{proc}, we recover (for the finite-dimensional case) the \bf data processing inequality \rm
 \begin{equation}\label{muller}D(\mathcal E\rho\| \mathcal E\sigma)\le D(\rho\| \sigma)\end{equation} for any trace-nonincreasing positive linear map $\mathcal E$ and any psdh matrices $\rho$ and $\sigma$ such that $\tr\mathcal E\rho=\tr \rho$. Note that \emph{complete} positivity of $\mathcal E$ is not assumed.

 The inequality~\eqref{muller}, in this generality, was first proved by  A.\ M\"uller-Hermes and D.\ Reeb \cite{MR}. They also covered the infinite-dimensional case. Their approach was based on the work of S.\ Beigi~\cite{B} establishing the data processing inequality for sandwiched R\'enyi divergences, with respect to quantum channels (completely positive trace-preserving linear maps).

 %For a quantum measurement, the condition of equality is that for all $i$ and all affine combinations $A$, we should have    $ E_iA^+=0$ or $ E_iA^-=0$.

{\color{black}{
\begin{Pro}\label{equality}  Let $\rho$ and $\sigma$ be psdh matrices with $\im\rho\subseteq\im\sigma$. Let $\mathcal E$ be a trace-nonincreasing positive linear map such that $\tr\mathcal E\rho=\tr \rho$.
\begin{enumerate}
\item[(a)]
Equality holds in~\eqref{muller} if and only if %every linear combination $A$ of $\rho$ and $\sigma$ has  $\mathcal EA^+\mathcal EA^-=0$, and %either $\im\mathcal E\rho\not\subseteq \im\mathcal E\sigma$, or
 \begin{equation}\label{hyp}\tr^+\mathcal E (\rho-t\sigma)=\tr^+ (\rho-t\sigma)\end{equation} holds for all $t> 0$.  This is equivalent to saying that every linear combination $A$ of $\rho$ and $\sigma$ has  $\mathcal EA^+\mathcal EA^-=0$, and %either $\im\mathcal E\rho\not\subseteq \im\mathcal E\sigma$, or
 \begin{equation*}%\label{hypo}
 \tr\mathcal E (\rho-t\sigma)^+=\tr (\rho-t\sigma)^+\end{equation*} holds for all $t> 0$.

 %\ietm{(b)]  When $\mathcal E$ is trace-preserving, the condition of equality is that

\item[(b)] For a quantum measurement, the condition of equality is that %{\color{black}{either there is an $i$ with $E_i\sigma=0\ne E_i\rho$, or}}
 for all $i$ and for all linear  combinations $A$ of $\rho$ and $\sigma$, we should have    $ E_iA^+=0$ or $ E_iA^-=0$.
\end{enumerate}
\end{Pro}

\begin{proof}
(a) First claim:  From Theorem~\ref{relent}, we see that equality holds in~\eqref{muller} if and only if  every affine combination $A=A(t)=(1-t)\rho+t\sigma$ of $\rho$ and $\sigma$ has  $\tr^\pm\mathcal EA(t)=\tr^\pm A(t)$   whenever $\pm t<0$.  This with the $+$ or the $-$ sign is equivalent to \eqref{hyp}  for $0<t<1$ and $t>1$, respectively.  For $t=1$, it follows by continuity.

The second claim is clear from  Lemma~\ref{proc}(b). %$\mathcal EA^+\mathcal EA^-=0$, and $\tr\mathcal E A(t)^\pm=\tr A(t)^\pm$ whenever $\pm t<0$.  In the first condition, we can clearly replace affine combinations by linear ombinations.

(b) Since quantum measurements are trace-preserving, the claim is immediate from (a). %ondition of equality is that {\color{black}{either there is an $i$ with $E_i\sigma=0\ne E_i\rho$, or}} for all $i$ and all affine combinations $A$, we should have    $ E_iA^+=0$ or $ E_iA^-=0$.
\end{proof}}}

%{\color{black}{We see from Theorem~\ref{relent} and from  Lemma~\ref{proc}(b) that}} equality holds in~\eqref{muller} if and only if every affine combination $$A=A(t)=(1-t)\rho+t\sigma$$ of $\rho$ and $\sigma$ has  $\mathcal EA^+\mathcal EA^-=0$, and $\tr\mathcal E A(t)^\pm=\tr A(t)^\pm$ whenever $\pm t<0$.
{\color{black}{
 In a  recent preprint~\cite{J} discussing the relevance of Theorem~\ref{relent} to the sufficiency of quantum channels and to hypothesis testing, it is pointed out  that when $\mathcal E$ is trace-preserving, the condition for equality in~\eqref{muller}  is~\eqref{hyp}, and that it can be reformulated as $\|\mathcal E(\rho-t\sigma)\|_1= \|\rho-t\sigma\|_1$ for all $t\ge 0$.
 When $\mathcal E$ is \emph{completely} positive and trace-preserving, it was known~\cite{BNPV, TV} that this preservation of the trace norm of linear combinations is equivalent to  the recoverability of $\rho$ and $\sigma$ with respect to $\mathcal E$. %the preservation of the quantum relative entropy.
  When $\mathcal E$ is 2-positive and trace-preserving, it was known~\cite{HM, P86, P88} that the    preservation of the quantum relative entropy is equivalent to %the preservation of $f$-divergences, and also is equivalent to numerous versions of
   the recoverability.}} % of $\rho$ and $\sigma$ with respect to $\mathcal E$. }}

%When $\mathcal E$ is trace-preserving, %we see
%from Proposition~\ref{equality} and Remark~\ref{equivalence}, or directly  that
% equality in~\eqref{muller} holds if and only if $\|\mathcal EA\|_1=\|A\|_1$ for all linear combinations $A$ of $\rho$ and $\sigma$.
%}}

 %The data-processing inequality~\eqref{muller} is known:have shown (in a more general setting)  that  quantum relative entropy satisfies this monotonicity.  %When $\mathcal E$ is a quantum channel (i.e., completely positive), it was already proved by S.\ Beigi~\cite{B}.

\subsection{The Holevo bound}  In \cite{H}, A.\ S.\ Holevo used his explicit formula for $S''$ to prove his celebrated upper bound on the quantity of information transmitted by a quantum communication channel.  We shall now show  how Theorem~\ref{main} quickly leads to a generalization of the same bound, which, however, also follows from \eqref{muller}.

Let $\mathcal E$ be a trace-nonincreasing positive linear map.
From Theorem \ref{main} and  Lemma~\ref{proc}, we see
that   $$
-{\color{black}{\left.S^{(m)}(\mathcal E\rho+t\mathcal E\sigma)\right|_{t=0}}}
\le -{\color{black}{\left.S^{(m)}(\rho+t\sigma)\right|_{t=0}}}%S^{(m)}(\rho+t \sigma)(0)
$$  for any psdh matrix $\rho$, any %traceless
 Hermitian matrix $\sigma$ satisfying $\im\sigma\subseteq\im\rho$, and any even $m\ge 2$. We have equality if and only if  every  combination $$A=A(t)=\rho+t\sigma$$  has  $\mathcal EA^+\mathcal EA^-=0$ and $\tr\mathcal E A^-=\tr A^-$. %$$\mathcal E((\rho+t\sigma)^+)\mathcal E((\rho+t\sigma)^-)=0$$ for all $t$. For a quantum measurement, the condition of equality is that there be no $i$ and $t$  with $\tr E_i(\rho+t\sigma)^\pm>0$ for both signs. % These  are known inequalities: A.\ M\"uller-Hermes and D.\ Reeb \cite{MR} have shown (in a more general setting)  that  quantum relative entropy satisfies this monotonicity, and the $S''$  statement easily follows.
For a quantum measurement $\mathcal E$, %a density matrix $\rho$ and a traceless Hermitian matrix $\sigma$,
 the condition of equality is that for all $i$ and $t$  we should have    $ E_iA(t)^+=0$ or $ E_iA(t)^-=0$. %there be no $i$ and  $t$   with $\tr E_iA(t)^\pm>0$ for both signs.

 In particular, $S-S\circ \mathcal E$  is a concave function on psdh matrices.

 {\color{black}{For psdh matrices $\rho_1$, \dots, $\rho_l$ and nonnegative weights $q_1$, \dots, $q_l$ summing to 1, let $\bar\rho=\sum_{j=1}^l q_j\rho_j$.}} Define the \emph{Holevo quantity} \begin{equation}\label{chidef}\chi (\rho_1, \dots, \rho_l;q_1, \dots, q_l) :=   S\left({\color{black}{\bar\rho}}\right)-\sum_{j=1}^lq_jS(\mathcal \rho_j){\color{black}{=\sum_{j=1}^lq_jD(\rho_j\|\bar\rho)}}.\end{equation}  From Theorem~\ref{main}(b), {\color{black}{or from Theorem~\ref{relent} together with the fact~\eqref{convex},}} we recover the well-known fact  that the Holevo quantity is nonnegative and convex as a function of $(\rho_1, \dots, \rho_l)$.

    By Jensen's inequality, for  any psdh matrices  $\rho_1$, \dots, $\rho_l$, and any weights  $q_1, \dots, q_l>0 $ summing to 1, we have \begin{equation}\label{holevo}\chi(\mathcal E\rho_1, \dots, \mathcal E\rho_l;q_1, \dots, q_l)\le \chi(\rho_1, \dots, \rho_l;q_1, \dots, q_l),\end{equation} with equality if and only if $\mathcal EA^+\mathcal EA^-=0$   and $\tr\mathcal E A^-=\tr A^-$ for every affine combination $A$ of $\rho_1$, \dots, $\rho_l$.

   In words:    the Holevo quantity is non-increasing under  trace-{\color{black}nonincreasing} positive linear maps. Note that complete positivity of the map  need not be assumed.

    When $\mathcal E$ is a quantum measurement, and each $\rho_j$ has trace 1, \eqref{holevo} is Holevo's inequality. The left hand side is the \emph{mutual  information} between the random input $j$  (whose distribution is given by the probabilities $q_j$) and the measurement output $i$  (whose conditional distribution is given by the conditional probabilities $\tr E_i\rho_j$ once $j$ has occurred).
 We have equality in~\eqref{holevo} if and only if for all  $i$ and all affine combinations $A$ of the $\rho_j$, we have % for all $i$ and all affine combinations $A$, we should have
     $ E_iA^+=0$ or $ E_iA^-=0$. %with $\tr E_iA^\pm>0$ for both signs.
%\newline
%\bf Holevo inequality \rm  \begin{align*}I:=S
%\left(\sum_{j=1}^l q_j\mathcal  E\rho_j\right)-\sum_{j=1}^lq_jS(\mathcal E\rho_j)\le \\ \le
%\chi (\rho_1, \dots, \rho_l;q_1, \dots. q_l) :=   S\left(\sum_{j=1}^l q_j\mathcal  \rho_j\right)-\sum_{j=1}^lq_jS(\mathcal \rho_j).\end{align*}
%Psdh matrices are taken to psdh matrices, therefore density matrices are taken to density matrices.
%\begin{Lemma}$\tr^{\pm}\mathcal EA\le \tr^{\pm}A$  for all Hermitian $A$.  Equality holds if and only if there is no $i$  with $\tr E_iA^\pm>0$ for both signs.
%\end{Lemma}
%Here $I$ is , and $\chi$ is the \emph{Holevo quantity}.

%It is worth noting that the above argument actually proves something stronger:

\section{Lower bounds on generalized divergences}\label{lowerbounds}
{\color{black}{This section is only loosely related to the rest of the paper. A simple application of the data processing inequality is presented.

 The most basic metric on quantum states  is given by the trace distance. It is therefore desirable to find inequalities that compare other measures}} {\color{black}{ of dissimilarity of quantum states, i.e.\ quantum divergences,  to the trace distance. Among the many divergences commonly studied, the \emph{quantum Jensen--Shannon divergence} $\chi(\rho_0, \rho_1; 1/2, 1/2)$, which is the Holevo quantity (concavity of the von Neumann entropy) with equal weights 1/2 and 1/2, has special interest because its square root is a metric~\cite{V}. }}

It was shown by F.\ Hiai, M.\ Ohya, and M.\ Tsukada \cite{HOC} that the minimum of the quantum relative entropy for two quantum states with prescribed trace distance is attained on binary classical states.  In this section, we shall use their method prove the analogous result for  any %`divergence-like'
quantity that depends on two quantum states and  is non-increasing under  quantum measurements. As examples of such quantities, we have already discussed in Section~\ref{dataprocess} the quantum relative entropy and the concavity of von Neumann entropy, but  the sandwiched R\'enyi divergence with parameter $\alpha>1$ is also  non-increasing, not just under quantum measurements, but under quantum channels (completely positive trace-preserving maps),  as was shown by S.\ Beigi \cite{B}.  More generally, M.\ M.\ Wilde~\cite{W} proved the same for optimized quantum $f$-divergences. An alternative proof  (with respect to  trace-preserving positive linear maps  satisfying a  certain Schwarz-type inequality)  was given by H.\ Li~\cite{Li}.%  --- and   the concavity of von Neumann entropy.

%For a simple application of the Holevo bound, l
Let  $\rho_0$ and $\rho_1$ be %distinct
 density matrices of dimension $\ge 2$. Set $$\rho_1-\rho_0=\|\rho_1-\rho_0\|_1\sigma,$$ where   $\sigma$ is traceless Hermitian  with 1-norm 1.  Let $\C^n=V_+\oplus V_-$ be an orthogonal decomposition with  $\sigma V_\pm\subseteq V_\pm$ and $\pm\sigma\ge 0$ on $V_\pm$. Let $E_\pm$ be the orthogonal projection onto $V_\pm$, and let $\mathcal E$ be the quantum measurement given by these two projections. Then, for any
density matrix $\rho$, we have  $\mathcal E\rho=\diag(\tr E_+\rho, \tr E_-\rho)$, whence $$\|\mathcal  E\rho_1-\mathcal E\rho_0\|_1=2\tr E_+(\rho_1-\rho_0)=2\|\rho_1-\rho_0\|_1\tr E_+\sigma=\|\rho_1-\rho_0\|_1.$$
We have $D(\mathcal E\rho_0\|\mathcal E\rho_1)\le D(\rho_0\|\rho_1)$ and \begin{equation}\label{chi}\chi(\mathcal E\rho_0,\mathcal E\rho_1;q_0, q_1)\le \chi(\rho_0,\rho_1;q_0, q_1)\end{equation}  for any positive $q_0$ and $q_1$ summing to 1. In both of these inequalities,
equality holds if and only if the density matrices $\rho_0$, $\rho_1$ and $|\sigma|$  are linearly dependent, i.e., lie on a line. Unique such $\rho_0$ and $\rho_1$ exist  for any prescribed  $\sigma$ and any prescribed values  of $0\le \tr E_+\rho_0\le\tr E_+\rho_1\le 1$.

More generally, let $\Delta$ be any function depending on two density matrices and satisfying the data processing inequality $\Delta(\mathcal E\rho_0\| \mathcal E\rho_1)\le \Delta(\rho_0\| \rho_1)$ for any quantum measurement $\mathcal E$ with two possible outcomes $+$ and $-$. Then we arrive at
\begin{Th}  For any quantum states (density matrices) $\rho_0,\rho_1\in M_n(\C)$, there exist binary classical states (diagonal 2-square density matrices)  $\rho_0'$ and $\rho_1'$  such that $\|\rho_1'-\rho_0'\|_1=\|\rho_1-\rho_0\|_1$ and $\Delta(\rho_0'\| \rho_1')\le\Delta(\rho_0\|\rho_1)$.
\end{Th}

For the case of the Holevo quantity, with the notations above,  we have
$S(\mathcal E\rho)=h(\tr E_+\rho)$, where \begin{equation}\label{h}h(x)=-x\log x-(1-x)\log(1-x)\end{equation} is the binary entropy function. Therefore, the mutual information is  $$\chi(\mathcal E\rho_0, \mathcal E\rho_1;q_0, q_1)=I(t_0, t_1;q_0, q_1):=h(q_0t_0+q_1t_1)-q_0h(t_0)-q_1h(t_1),$$   where  $t_j=\tr E_+\rho_j$. Observe that $$t_1-t_0=\tr E_+(\rho_1-\rho_0)=\|\rho_1-\rho_0\|_1\tr E_+\sigma=\|\rho_1-\rho_0\|_1/2.$$
We arrive at
\begin{Th}\label{lower}For any nonnegative $q_0$ and $q_1$ summing to $1$, we have
%\begin{enumerate}\item[(a)]
\begin{align*}
\chi(\rho_0, \rho_1; q_0, q_1)\ge \\\ge% I(\tr E_+\rho_0, \tr E_+\rho_1;q_0, q_1)\ge\\\ge
\min \{I(t_0, t_1;q_0, q_1):0\le t_0\le t_1\le 1,  t_1-t_0=\|\rho_1-\rho_0\|_1/2.\}\end{align*}
 % In the first inequality, equality holds if and only if the density matrices $\rho_0$, $\rho_1$ and $|\sigma|$  are linearly dependent, i.e., lie on a line. Unique such $\rho_0$ and $\rho_1$ exist  for any prescribed  $\sigma$ and any prescribed values  of $0\le \tr E_+\rho_0< \tr E_+\rho_1\le 1$.
\end{Th}
This theorem and the possibility  of equality in \eqref{chi} tell us that for the Holevo quantity, or `quantum entropy concavity' $\chi(\rho_0, \rho_1; q_0, q_1) $,  the largest  lower bound that depends only on $\rho_1-\rho_0$ and $q_1$ is  the `minimal classical binary entropy concavity', i.e., the minimum in Theorem~\ref{lower}. It does not seem possible to compute this minimum exactly. There are various ways to get weaker but more explicit lower bounds. A simple way is to use the convexity and symmetry of $-h''(x)=1/x+1/(1-x)$ to  prove that the minimum is $$\ge
4q_0q_1
\left(
h\left(\frac12\right)
-h\left(\frac{2+\|\rho_1-\rho_0\|_1}4\right)
\right),$$ with equality if and only if $q_1=1/2$ or $\rho_0=\rho_1$.
 Note that $h(1/2)=\log 2$. For $\rho_0\ne\rho_1$ and $q_0q_1>0$, this weaker lower bound on $$\chi(\rho_0, \rho_1; q_0, q_1) $$ is still strictly  greater than the previously known   lower bound  $$q_0q_1\|\rho_1-\rho_0\|_1^2/2$$ due to I.\ H.\  Kim \cite{K}.  This is because $-h''$  is minimal, with value 4, only at $1/2$.

 For lower bounds depending on other parameters of $\rho_0$ and $\rho_1$, and also for upper bounds, see \cite{KR}.  %; they are equal only if $\rho_0=\rho_1$.

 {\color{black}{\section{General probabilistic theory}\label{GPT}
 In this vast generalization of quantum theory~\cite{Pl}, the set of density matrices is replaced by a  more general \emph{state space}, i.e.,
 a convex body $K$ in a finite dimensional real affine space $\mathbb A$. We may assume that $K$ spans $\mathbb A$ as an affine space. Then points of $\mathbb A$, called \emph{virtual states}, play the role of Hermitian matrices with trace 1. The vectors of the underlying vector space $V=\mathbb A-\mathbb A$ play the role of traceless Hermitian matrices. %will be called \emph{virtual states}.
  Let $E$ be the set of \emph{effects}, i.e., affine  functions $e:\mathbb A\to\R$ such that $0\le e\le 1$ on  $K$. For a virtual state $A\in \mathbb A$, we define
\[\tr^+A=\max_{e\in E} e(A)\] and
\[\tr^-A=-\min_{e\in E} e(A).\]
Clearly, $\tr^+-\tr^-=1$.  For a  state $\rho\in K$, we have $\tr^+\rho=1$ and $\tr^-\rho=0$.

For two (virtual) states $\rho_0$ and  $\rho_1$, we define their \emph{trace distance} to be  $$\|\rho_1-\rho_0\|_1=%\tr^+A+\tr^-A=
2\max_{e\in E}\left(e(\rho_1)-e(\rho_0)\right).$$ }}%where $A=\rho_1-\rho_0$.

{\color{black}{
  The role of quantum measurement is played by  a \emph{general measurement}, or \emph{partition of unity}, i.e., a  sequence $e_1, \dots, e_k\in E$ of effects such that $e_1+\dots +e_k=1$  (the constant 1 function). More generally, the role of a}} {\color{black}{ positive trace-preserving map is played by an affine map $\mathcal E:\mathbb A\to \mathbb A'$ such that  $\mathcal E(K)\subseteq K'$, where $K'$ is another state space spanning another affine space $\mathbb A'$. Since $\mathcal E$ is determined by its restriction to $K$, we will refer to it simply as an affine map $\mathcal E:K\to K'$. When $\mathcal E$ is a general measurement, $K'$ is the set of $k$-ary classical states, i.e., the  simplex    with $k$ vertices.  The role of Lemma~\ref{proc}(a) is played by
  \begin{Lemma}\label{elemigeneral} For  any affine map $\mathcal E:K\to K'$,  %such that  $\mathcal E(K)\subseteq K'$,
   we have $\tr^\pm\mathcal EA\le \tr^\pm A$ for any virtual state  $A\in\mathbb A$.
  \end{Lemma}
  \begin{proof} For any effect $e'\in E'$, we have a corresponding effect  $e=e'\circ\mathcal E\in E$ such that $e(A)= e'(\mathcal EA)$.
  \end{proof}
}}

{\color{black}{
For two states $\rho,\sigma\in K$, we define the \emph{general relative entropy} $D(\rho\|\sigma)$ to be the integral~\eqref{relentformula}.  From  Lemma~\ref{elemigeneral}, we get
\begin{Th}The data processing inequality~\eqref{muller} holds for any affine map $\mathcal E:K\to K'$. %  such that  $\mathcal E(K)\subseteq K'$.
\end{Th}

For states $\rho_1, \dots, \rho_l\in K$ and nonnegative weights $q_1$, \dots, $q_l$ summing to 1, we define the \emph{general Holevo quantity} $\chi(\rho_1, \dots, \rho_l;q_1, \dots, q_l)$  to be the last sum in~\eqref{chidef}. From the previous Theorem, we get
\begin{Th}\label{generalholevo} The inequality~\eqref{holevo}  holds for any affine map $\mathcal E:K\to K'$; %  such that  $\mathcal E(K)\subseteq K'$;
 in particular, for any general measurement.
\end{Th} This is   an extension of Holevo's inequality to general probabilistic theory.

As an example, let $K=\{\rho\in\R^d:|\rho|\le 1\}$ be the unit ball. The role of von Neumann entropy is played by the function $S(\rho)=h((1-|\rho|)/2)$, where $h$ is the binary entropy function~\eqref{h}.  For any given $\sigma$ in the interior of the ball  $K$, the sum $S(\rho)+D(\rho\|\sigma)$ is an affine function of $\rho$. Indeed, for any linear  subspace $W\le  \R^d$ of dimension three (or less)  containing $\sigma$, we may identify $K\cap W$ with the Bloch ball of 2-square density matrices (or a central section of it), and then, for all $\rho\in K\cap W$,  the above sum becomes $-\tr\rho\log\sigma$.  Therefore, {\color{black} for any states $\rho_1, \dots, \rho_l\in K$ and nonnegative weights $q_1$, \dots, $q_l$ summing to 1,  we have \[S(\bar\rho)=S(\bar\rho)+D(\bar\rho\|\bar\rho)=\sum_{j=1}^lq_j(S(\rho_j)+D(\rho_j\|\bar\rho))%=\sum_{j=1}^lq_jS(\rho_j)+\sum_{j=1}^l q_jD(\rho_j\|\bar\rho)
,\] where $\bar\rho=\sum_{j=1}^lq_j\rho_j$. This means that}
the identity~\eqref{chidef} holds in this setting, providing an explicit form of the  Holevo quantity $\chi$ in terms of the entropy function $S$, just as in quantum information theory.

For general state spaces, however, $\chi$ seems unlikely to have a  computable closed form, and therefore Theorem~\ref{generalholevo} will be more difficult to apply than in the quantum case. However, exploring the properties of $D$ and $\chi$ in this framework might be an interesting topic of future research.

The results of Section~\ref{lowerbounds} also have straightforward extensions to the general probabilistic setting. Indeed, for any states $\rho_0, \rho_1\in K$, there is an effect $e:K\to [0,1]$ such that \[e(\rho_1-\rho_0)=\|\rho_1-\rho_0\|_1/2.\] Then $\mathcal E=(e,1-e)$ is a general measurement such that the binary classical states $\mathcal E\rho_0$  and $\mathcal E\rho_1$ satisfy  $$\|\mathcal E\rho_1-\mathcal E\rho_0\|_1=\|\rho_1-\rho_0\|_1.$$  Therefore, we have
\begin{Th}If the trace distance is given, then any general divergence satisfying the data processing inequality for two-part general measurements (for example, the general relative entropy, or the general Holevo quantity for two states and two given weights) will have the same infimum on pairs of general states as on  pairs of  binary classical states.
\end{Th}

\section{R\'enyi  divergence}\label{renyi} Throughout this section, let $1\ne \alpha>0$.  For $x,y\in\R_{\ge 0}^n$ with $\sum_{i=1}^nx_i=1$, recall the definition of R\'enyi  divergence of order $\alpha$:
\[D_\alpha(x\|y)=%\begin{cases}
\frac1{\alpha-1}\log{\color{black}\sum_{i=1}^n}x_i^\alpha y_i^{1-\alpha},
\]
{\color{black} where, in case of $\alpha>1$, we define $0^{1-\alpha}=\infty$.  We define $0\cdot\infty =0$, $\log \infty =\infty$ and $\log 0=-\infty$. Thus, $D_\alpha(x\|y)$ will be finite if and only}
if $\alpha>1$  and  $x_i=0$  whenever $y_i=0$, or if $\alpha<1$ and  $x_iy_i>0$
 for some $ i$. Otherwise, we {\color{black} have} $D_\alpha(x\|y)=\infty$.
%\end{cases}\]

 It can be checked that {\color{black} \begin{equation}\label{renyitag}x_i^\alpha y_i^{1-\alpha}=x_i+(\alpha-1)\left(x_i-y_i+\alpha\int_{-\infty}^\infty\frac{|t|^{\alpha-2}dt}{|t-1|^{\alpha+1}}((1-t)x_i+ty_i)^-\right),\end{equation}  where $\lambda^-=-\min(\lambda, 0)$. %The integral is $\infty$ if and only if $\alpha>1$ and $x_i\ne 0=y_i$. If $x_i=y_i=0$, then  both sides of \eqref{renyitag} are zero.
 After summing \eqref{renyitag} for all $i$, we get that} $D_\alpha(x\|y)$ is equal to \begin{equation}\label{renyiformula}\frac1{\alpha-1}\log\left(1+(\alpha-1)\left({\color{black}1-\tr\sigma}+\alpha\int_{-\infty}^\infty\frac{|t|^{\alpha-2}dt}{|t-1|^{\alpha+1}}\tr^-A(t)%()
 \right)\right),\end{equation} where $\rho=\diag(x_1, \dots, x_n)$, $\sigma=\diag(y_1, \dots y_n)$, and $A(t)=(1-t)\rho+t\sigma$.  {\color{black} This fact %is essentially known; it
  is closely related to \cite[Proposition 3]{SV}.}

  Let us now allow $\rho\in M_n(\C)$ to be any density matrix, and $\sigma\in M_n(\C)$ to be any psdh matrix. {\color{black} Let us first study the sum in the inner large parentheses in \eqref{renyiformula}. It is $\ge1-\tr\sigma>-\infty$.}   Using the identity~\eqref{atiras} as in  the proof of Theorem~\ref{relent}, {\color{black} it} can be rewritten as \begin{equation}\label{parenth}\alpha\left(\int_{-\infty}^0\frac{|t|^{\alpha-2}dt}{(|t|+1)^{\alpha+1}}\left(\tr^+A(t)-1%\tr\rho
 \right)+\int_1^\infty \frac{t^{\alpha-2}dt}{(t-1)^{\alpha+1}}\tr^-A(t)\right).\end{equation}

 When $\alpha>1$,  observe  that $\tr^\pm A(t)\ge 0$ for all $t$; moreover, we have  $\tr^+A(0)=1$, %and $\tr^+A(t)$  is continuous,
  so $\tr^+A(t)>0$ for $t<0$ close enough to $0$.  Therefore, \eqref{parenth} is \[>-\alpha\int_{-\infty}^0\frac{|t|^{\alpha-2}dt}{(|t|+1)^{\alpha+1}}=-\frac1{\alpha-1}.\] {\color{black}Still assuming $\alpha>1$, observe also that \eqref{parenth} is $\infty$ if and only if $\im\rho\not\subseteq\im\sigma$.}

When $\alpha<1$, observe that $\tr^+A(t)\le 1-t=|t|+1$ for $t<0$ and $\tr^-A(t)\le t-1$ for $t>1$, with equality if and only if $\im\rho\subseteq\ker\sigma$, whence \eqref{parenth} is \[\le\alpha\left(\int_{-\infty}^0\frac{|t|^{\alpha-1}dt}{(|t|+1)^{\alpha+1}}+\int_1^\infty \frac{t^{\alpha-2}dt}{(t-1)^{\alpha}}\right)=\frac1{1-\alpha} ,\]  with the same condition for equality.

{\color{black}We deduce}
 \begin{Pro}\label{logarg} For any density matrix $\rho$ and psdh matrix $\sigma$, the expression under  the logarithm in~\eqref{renyiformula} is nonnegative. It is zero if and only if  $\alpha<1$ and $\im\rho\subseteq\ker\sigma$. {\color{black} It is $\infty$ if and only if $\alpha>1$ and $\im\rho\not\subseteq\im\sigma$.}
 \end{Pro}

 %the right hand side of ~
 Thus, \eqref{renyiformula} defines a quantum generalization {\color{black}$\Delta_\alpha(\rho\| \sigma)$} of R\'enyi divergence which is unitarily invariant {\color{black} since semidefiniteness, $\tr$, and $\tr^-$ are all unitarily invariant. It takes values in $(-\infty, \infty]$. It is finite if and only if $\alpha>1$ and $\im\rho\subseteq\im\sigma$, or $\alpha<1$ and $\im\rho\not\subseteq\ker\sigma$.
 When $\tr\sigma\le 1$, we have  $\Delta_\alpha(\rho\|\sigma)\ge 0$, with equality if and only if $\rho=\sigma$. %{\color{black}The divergence $\Delta_\alpha(\rho\| \sigma)$}

 \begin{Pro}\label{cont} For any fixed density matrix $\rho$ and psdh matrix $\sigma$, the divergence $\Delta_\alpha(\rho\| \sigma)$ tends to Umegaki's quantum relative entropy $D(\rho\|\sigma)$ as $\alpha\to 1$.
 \end{Pro}

 \begin{proof} %When $\im\rho\subseteq\ker\sigma$, we have $\Delta_\alpha(\rho\|\sigma)=\infty=D(\rho\|\sigma)$.
 By the Monotone Convergence Theorem, both integrals in~\eqref{parenth} converge to their respective  values at $\alpha=1$ as $\alpha\to 1$. Due to Theorem~\ref{relent}, the value of \eqref{parenth} for $\alpha=1$ is $D(\rho\|\sigma)$. If this value is finite, then \eqref{parenth} multiplied by $\alpha-1$   converges to zero as $\alpha\to 1$, whence $\Delta_\alpha(\rho\|\sigma)\to D(\rho\|\sigma)$ as claimed.  If $D(\rho\|\sigma)=\infty$, then, %$\im\rho\not\subseteq\im\sugma$, whence, for all $\alpha>1$, we have $\Delta_\alpha(\rho\|\sigma)=\infty)$, and,
   for all finite $C>0$, we have  \[\liminf_{\alpha\to 1} \Delta_\alpha(\rho\|\sigma)\ge\lim_{\alpha\to 1}\frac1{\alpha-1}\log(1+C(\alpha-1))=C,\] whence $\Delta_\alpha(\rho\|\sigma)\to\infty$ as $\alpha\to 1$.
 \end{proof}
 }

 Due to Lemma~\ref{proc}(a) {\color{black}and formula~\eqref{parenth}},  the data processing inequality {\color{black} $\Delta_\alpha(\mathcal E\rho\|\mathcal E\sigma)\le\Delta_\alpha(\rho\|\sigma)$ holds} for trace-nonincreasing positive linear maps $\mathcal E$ such that $\mathcal E\rho$ is again a density matrix.
 Exploring the further properties of {\color{black} $\Delta_\alpha$}, and relating it to existing quantum generalizations of R\'enyi divergence, might be an interesting topic for future research.\footnote{\color{black}{
 \it Note added in Version 4 of this paper:
\rm Such `future research', relating $\Delta_\alpha$ to Petz--R\'enyi and sandwiched R\'enyi divergences, was carried out and posted  in a very recent preprint by Hirche and Tomamichel \cite[Section 3]{HT}. Also, a quantitative version of Proposition~\ref{cont} is  given in \cite[Subsection 3.2]{HT}.}}

  Note that %the right hand side of ~
 \eqref{renyiformula} allows us to define a notion of R\'enyi  divergence  in the general probabilistic setting of Section~\ref{GPT}.  For any states $\rho, \sigma\in K$, we put\[\Delta_\alpha(\rho\| \sigma)=\frac1{\alpha-1}\log\left(1+(\alpha-1)\alpha\int_{-\infty}^\infty\frac{|t|^{\alpha-2}dt}{|t-1|^{\alpha+1}}\tr^-((1-t)\rho+t\sigma)%()
 \right).\] Similarly to Proposition~\ref{logarg}, it is easy to check that the expression under the logarithm is nonnegative even if $\alpha<1$. Indeed, we have $\tr^-A(t)\le |t|$ for $t<0$ and $\tr^-A(t)\le t-1$ for $t>1$.

 Obviously $\Delta_\alpha(\rho\|\sigma)\ge 0$, with equality if and only if $\rho=\sigma$. %When $\alpha<1$, it is $\in [0,1]$, where the nonnegativity is checked similarly to Proposition~\ref{logarg}. %where  $A(t)=(1-t)\rho+t\sigma$.
 The data processing inequality remains valid:
 \begin{Th}
 For any affine map $\mathcal E:K\to K'$, we have \[\Delta_\alpha(\mathcal E\rho\|\mathcal E\sigma)\le\Delta_\alpha(\rho\|\sigma).\]
 \end{Th}

 \begin{proof} This is clear from Lemma~\ref{elemigeneral}.
 \end{proof}
}}

\end{document}